\documentclass[aps,prd,groupedaddress,superscriptaddress,11pt,nofootinbib,preprintnumbers,floatfix]{revtex4-2}

\usepackage{dsfont}
\usepackage{natbib}
\usepackage{url}
\usepackage{epstopdf}
\usepackage{color}
\usepackage{fullpage}
\usepackage[top=2cm, bottom=4.5cm, left=2.5cm, right=2.5cm]{geometry}
\usepackage{amsmath,amsthm,amsfonts,amssymb,amscd}
\usepackage{lastpage}
\usepackage{enumerate}
\usepackage[shortlabels]{enumitem}
\usepackage{fancyhdr}
\usepackage{mathrsfs}
\usepackage{xcolor}
\usepackage{graphicx}
\usepackage{listings}
\usepackage{hyperref}

\hypersetup{%
  colorlinks=true,
  linkcolor=blue,
  linkbordercolor={0 0 1}
}

\usepackage{subcaption}

\newcommand{\be}{\begin{equation}}
\newcommand{\ee}{\end{equation}}
\newcommand{\bea}{\begin{eqnarray}}
\newcommand{\eea}{\end{eqnarray}}
\newcommand{\nn}{\nonumber}

\newcommand{\sq}{\sqrt{q}}
\renewcommand{\H}{\mathcal{H}}

\begin{document}

\title{Monte Carlo quantum cosmology with a modified Euclidean action}

\bigskip

\author{Malaik Kabir}
\email{malaikkabir@gmail.com}
\email{24100156@lums.edu.pk}
\affiliation{Department of Physics, Syed Babar Ali School of Science and Engineering, Lahore University of Management Sciences, Lahore 54792, Pakistan} 

\author{Syed Moeez Hassan}
\email{syed\_hassan@lums.edu.pk}
\affiliation{Department of Physics, Syed Babar Ali School of Science and Engineering, Lahore University of Management Sciences, Lahore 54792, Pakistan} 

\date{\today}

\begin{abstract}

\vskip 1cm

We numerically study the Euclidean quantum cosmology of a closed, homogeneous and isotropic universe with a cosmological constant. A dust field acts as a clock, and we compute the ground state wavefunction, correlation function, and mean volume of the universe by performing path integral Monte Carlo simulations. To make the path integral convergent, the argument of the exponent in the weight of the path integral is chosen to be either the absolute value of the Euclidean action, or its square. We compare our results to the standard case (with a hard cutoff at zero action), and find that the absolute value produces similar results, whereas the squared method gives substantially different results. We detail two different methods of `fixing' the squared case, and test them for the simple harmonic oscillator, and for this cosmological model. We also find that for a large positive cosmological constant, the universe undergoes multiple cycles of expansion and contraction for all three weights, and reaches much larger sizes with the squared weight.

\end{abstract}

\maketitle
\vskip 0.5cm
\section{Introduction}

Numerical techniques are increasingly becoming important in quantum gravity, where models are sufficiently complex that they preclude a full analytical treatment. Among these techniques, Monte Carlo methods are particularly noteworthy and have been an important tool in physics for many decades, with perhaps the most famous instance being Lattice Quantum Chromo-Dynamics (QCD) \cite{2010LNP...788.....G, Ratti:2018ksb}. These techniques have also been applied to quantum gravitational models with Causal Dynamical Triangulations (CDT) \cite{Ambjorn:2013tki}, causal sets \cite{Rideout:1999ub}, and spin foams \cite{Perez:2012wv} providing a few examples.

Here, we apply Markov Chain Monte Carlo (MCMC) techniques to evaluate the gravitational path integral for a homogeneous and isotropic cosmological spacetime. There are two major obstructions on this path: The first is the problem of time in quantum gravity \cite{Anderson:2017jij}, which for our cosmological model implies that the Hamiltonian is constrained to vanish. Therefore, to capture the correct physics, this constraint has to be implemented in the path integral. The second is the unboundedness of the gravitational action from below \cite{Gibbons:1978ac, Marolf:1996gb, Horowitz:2025zpx}. The standard (Lorentzian) path integral is evaluated with the weight $\exp(iS)$ where $S$ is the action of the theory. This factor, however, is highly oscillatory, and therefore not suitable for a direct numerical computation. A standard way to resolve this is to perform a Wick rotation, which rotates `time' by 90 degrees about the imaginary axis, thereby converting the weight to $\exp(-S_E)$, where $S_E$ is known as the Euclidean action. For most ordinary systems, $S_E$ is bounded from below, and therefore, a convergent path integral exists. The gravitational Euclidean action, however, is unbounded from below. 

A way around the first problem is provided by using relational variables at the classical level, before proceeding to the quantum theory. In particular, the idea is to use a classical `clock' consisting of some degree(s) of freedom of the model, solve the Hamiltonian constraint (strongly) at a classical level to get to a reduced physical phase space, and then quantize the remaining degrees of freedom and study their dynamics with respect to the chosen clock. There are many different types of clocks that could be chosen (for example, \cite{Blyth:1975is}), and while classically, they result in the same dynamics, the quantum theories are not equivalent \cite{Malkiewicz:2014fja, Malkiewicz:2015fqa, Malkiewicz:2016hjr, Gielen:2021igw}. An interesting choice of clock is provided by a pressureless dust field, which renders the physical Hamiltonian a linear function of the remaining variables \cite{Husain:2011tk} (instead of being a square root as with other gravitational or matter clocks). 

There is no straightforward solution to the gravitational action being unbounded, and some mechanism has to be implemented `by hand'. In \cite{Ali:2018nql}, the idea was to introduce a lower bound at $S_E = 0$, and then study the resultant theory. Here, we propose two alternate methods that rely on modifying the weight/measure in the Euclidean path integral. In particular, we consider path integrals with weights $\exp(-|S_E|)$, and $\exp(-S_E^2)$, and study the resultant quantum theory. Beyond the obvious advantage that these weights provide a bounded path integral even when the Euclidean action $S_E$ is unbounded from below, they also ensure that the classical (Euclidean) theory is the same as the one arising from using the standard measure \cite{Modanese:2019bvz, Modanese:2020gna}. The methods of sampling paths with $S_E = 0$, as well as changing the measure to $\exp(-|S_E|)$ (along with some other approaches, for instance, assigning extremely low weightages to paths with $S_E < 0$, or introducing hard cut-offs in the potentials/domains to restrict the paths from going outside a bounded region), were also presented in \cite{Berger:1988rp, Berger:1989jm, Berger:1993fm}. The main difference however (along with exploring another path integral measure here), is that we fix a time gauge first (by solving the Hamiltonian constraint), and then quantize the reduced theory. This also means that we have an extra degree of freedom in our theory, which after fixing the clock, appears in the gravitational sector.

We present our model in Sec \ref{sec-model}, our main results are discussed in Sec \ref{sec-qc} (with Appendix \ref{sec-app} providing more details on how this method works for a well known system -- the simple harmonic oscillator), and we summarize in Sec \ref{sec-sum}. We work in $c = \hbar = 8\pi G = 1$ units throughout.

\section{The model} \label{sec-model}

We start with General Relativity written in the Arnowitt-Deser-Misner (ADM) form, coupled to a pressureless dust field $T$,
\be
\label{full-theory}
S = \int d^3x dt ~ [ \pi^{ab} \dot{q}_{ab} + p_T \dot{T} - N \H - N^a C_a],
\ee
where,
\bea
\label{full-ham}
\H &=& \H_G + \H_D, \nn \\
\H_G &=& \dfrac{1}{\sq} \Bigg(\pi^{ab} \pi_{ab} - \frac{1}{2} \pi^2 \Bigg) + \sq ( \Lambda - R), \nn \\
\H_D &=& p_T \sqrt{1 + q^{ab} \partial_a T \partial_b T }, \\
C &=& C^G_a + C^D_a, \nn \\
C^{G}_a &=& D_b \pi^b_a, \nn \\
C_a^D &=& -p_T \partial_a T,
\eea
$(q_{ab}, \pi^{ab})$ and $(T, p_T)$ are the gravitational and dust phase space variables respectively, $q$ is the determinant of the spatial metric $q_{ab}$, $\pi$ is the trace of $\pi^{ab}$, $R$ is the 3-Ricci scalar curvature, $\Lambda$ is the cosmological constant, $N$ is the lapse and $N^a$ is the shift.

We then reduce to a Friedmann-Lemaitre-Robertson-Walker (FLRW) universe with the ansatz,
\be
q_{ab} = \frac{3}{8} A^{4/3}(t) h_{ab} , ~ \pi^{ab} = 2 A^{-1/3}(t) p_A(t) \sqrt{h} h^{ab},
\ee
where, $h_{ab} = f(r) e_{ab}$ with $f(r) = 1 + kr^2/4$ and $e_{ab} = diag(1,1,1)$, $k$ is the spatial curvature of the universe with three distinct possibilities: $k=0$ (flat), $k>0$ (closed), and $k<0$ (open),  $A(t)$ is related to the usual scale factor $a(t)$ as $a(t) \sim A^{2/3}(t)$, and $p_A$ is the momentum canonically conjugate to $A$. This ansatz corresponds to writing the FLRW metric in the isotropic form,
\be
ds^2 = -N^2(t) dt^2 + \dfrac{a^2(t)}{1 + kr^2/4} \left(dr^2 + r^2 d\Omega^2 \right).
\ee
We also assume that the dust field is homogeneous (to ensure consistency with a homogeneous universe),
\be
T = T(t), ~ p_T = p_T(t).
\ee
With this homogeneity ansatz, the diffeomorphism constraint vanishes identically and the Hamiltonian constraint becomes,
\be
\label{Ham-constr-cosm}
H =  -\dfrac{1}{2} p_A^2  + \dfrac{1}{2} \Lambda A^2 - k A^{2/3} + p_T \approx 0
\ee
(here we have rescaled the constants as: $\Lambda \rightarrow 3 \Lambda/4$, and $k \rightarrow (3/8)^{1/3} k$).

We now proceed to use the dust field as a clock, explicitly solve the Hamiltonian constraint, and get to a reduced physical Hamiltonian \cite{Husain:2011tk, Husain:2011tm}. We first identify our time variable with the dust field: $t=-T$. The Hamiltonian constraint (\ref{Ham-constr-cosm}) is then (strongly) solved for the momentum conjugate to this choice of time $p_T$, and the \emph{manifestly positive} physical Hamiltonian becomes (this corresponds to the choice $N=-1$, we refer the reader to \cite{Hassan:2017cje} for details),
\be
\label{phys-ham}
H_p = p_T = \dfrac{1}{2} p_A^2  - \dfrac{1}{2} \Lambda A^2 + k A^{2/3},
\ee
with the gauge-fixed action,
\be
S = \int dt \left( \dfrac{1}{2} \dot{A}^2 + \dfrac{1}{2} \Lambda A^2 - k A^{2/3} \right). 
\ee
This form of the Hamiltonian clarifies the utility of the variable $A(t)$, as we note that the kinetic term is the standard kinetic term of a particle, and for $k=0$, we have either a free particle ($\Lambda = 0$), a simple harmonic oscillator ($\Lambda < 0$), or an inverted oscillator ($\Lambda > 0$). This makes studying the quantum theory relatively straightforward \cite{Ali:2018vmt, Maeda:2015fna}. The presence of a non-zero spatial curvature however, precludes an analytical treatment, and has to be studied via numerical methods \cite{Ali:2018nql}. In what follows, we study the quantum theory of the $k \neq 0$ case through the path integral approach which starts with the path integral,
\be
\int \mathcal{D}A(t) ~ e^{iS[A(t)]}.
\ee
The presence of `$i$' in the exponent makes this integral highly oscillatory and therefore, not suitable for a numerical study. The standard approach is to Wick rotate the time variable ($t \rightarrow -it$) which renders the exponent real, and makes the path integral,
\be
\int \mathcal{D}A(t) ~ e^{-S_E[A(t)]},
\ee
where, $S_E$ is the Euclidean action, which for our model takes the form,
\be
\label{euc-action}
S_E = \int_0^T dt \left( \dfrac{1}{2} \dot{A}^2 - \dfrac{1}{2} \Lambda A^2 + k A^{2/3} \right). 
\ee
The above Euclidean action is different from standard quantum mechanical actions in one very significant way: it is unbounded from below (when $\Lambda > 0$, $k<0$, or both). This implies that there is no stable ground state, and that the (Euclidean) path integral is not convergent. This issue is well known for gravitational theories. In \cite{Ali:2018nql}, this was overcome by restricting to the set of paths for which $S_E = 0$. This provides a fixed cutoff, and renders the Euclidean path integral convergent. Here, we explore two alternate possibilities which rely on modifying the exponential weight factor in the path integral. We consider path integrals of the form,
\be
\label{mod-pi}
\int \mathcal{D}A ~ e^{-|S_E|} ~~ \text{and} ~~ \int \mathcal{D}A ~ e^{- \beta S_E^2},
\ee
where $\beta$ is an arbitrary parameter.

\subsection{Markov Chain Monte Carlo}

We numerically evaluate the path integrals above using Markov Chain Monte Carlo (MCMC) techniques. In particular, we use the Metropolis-Hastings algorithm to sample paths. We detail our procedure below:

\begin{enumerate}

    \item Start with a random array (lattice) of $N$ entries where each entry is chosen according to a uniform random distribution in the range [$-x, +x$]. This will be our path.
    
    \item Take the first value $x_1$ in the array and generate a random value $x_{\text{new}}$ sampled from a uniform distribution in the range [$x_1 - \Delta$, $x_1 + \Delta$]. Here $\Delta$ is a parameter that controls the range within which new path values are generated.
    
    \item Compute the action $S_{\text{old}}$ for the path with $x_1$ as the first entry and the action $S_{\text{new}}$ for the path with $x_{\text{new}}$ as the first entry.
    
    \item If $S_{\text{new}} < S_{\text{old}}$ accept $x_{\text{new}}$ as the new entry and change the path accordingly. On the other hand if $S_{\text{new}} > S_{\text{old}}$ accept $x_{\text{new}}$ with probability $e^{-(S_{\text{new}} - S_{\text{old}})}$.
    
    \item Repeat steps 2 - 4 for all $N$ values of the path array. This completes one Monte Carlo step.
    
    \item Perform $N_{MC}$ number of Monte Carlo steps. This completes one simulation run.
    
\end{enumerate}

In order to minimize the dependence of the Markov Chain on the choice of a random initial path array, we run the chain for a certain number of iterations before counting them towards $N_{MC}$. Throughout, this portion of the Markov Chain, which is not included in the sampling, is referred to as the ``thermalization phase/step'' of the simulation and the number of iterations constituting this step is denoted by the variable $N_{\text{therm}}$. Additionally, in order to minimize correlations between successive iterations of the Markov Chain during a simulation, we only count every $\bar{n}$'th iteration and discard the intervening iterations.

In our MCMC calculations, the kinetic energy term in the action is calculated using the forward difference method. The final lattice point, therefore, does not contribute to the kinetic energy. The potential energy contribution, on the other hand, is from all lattice points. For a large enough lattice, this should not affect any physical results. Additionally, all lattice points -- except where it is specified that $A(0) = 0$ -- are updated on each iteration based on the Metropolis-Hastings algorithm.

We now turn to applying this method to our cosmological model in the next section. The implications of these modified path integrals (\ref{mod-pi}) for a standard system -- the harmonic oscillator, and finding a suitable value for the parameter $\beta$ are discussed in Appendix \ref{sec-app}.

\section{Quantum Cosmology} \label{sec-qc}

We begin with discretizing the action of our model (\ref{euc-action}) to get,
\be
S_E = \epsilon \sum_{i=1}^{N-1} \left( \dfrac{(A_{i+1} - A_i)^2}{{2 \epsilon}^2} - \frac{1}{2} \Lambda A_i^2 + k {A_i}^{2/3} \right),
\ee
where $\epsilon$ is the lattice spacing, and satisfies $T = N \epsilon$. We focus here on the case of a spatially closed universe, and fix $k = +1$. The value of the cosmological constant $\Lambda$ then controls whether the action is bounded or unbounded from below, and we explore each of these cases separately in the subsections below.

\subsection{Bounded case ($\Lambda = -1$)}

We start with the case where the Euclidean action is bounded from below ($\Lambda = -1$), and hence a unique ground state exists. We first calculate the analog of the Hartle-Hawking `no-boundary' ground state wavefunction \cite{Hartle:1983ai} by keeping $A(0) = 0$ fixed for all paths. Various other parameter values were chosen to be, $\epsilon = 0.01$, $N = 1500$ (i.e., $T=15$), $N_{MC} = 10^{6}$, $\Delta = 0.2$ and $\bar n = 5$ (the bin-sizes varied in each case but were defined such that there were 1000 bins within the relevant range for each simulation). The initial path was chosen to be in the range [-5, 5].

\begin{figure} 
    \centering
    \includegraphics[width=0.8\linewidth]{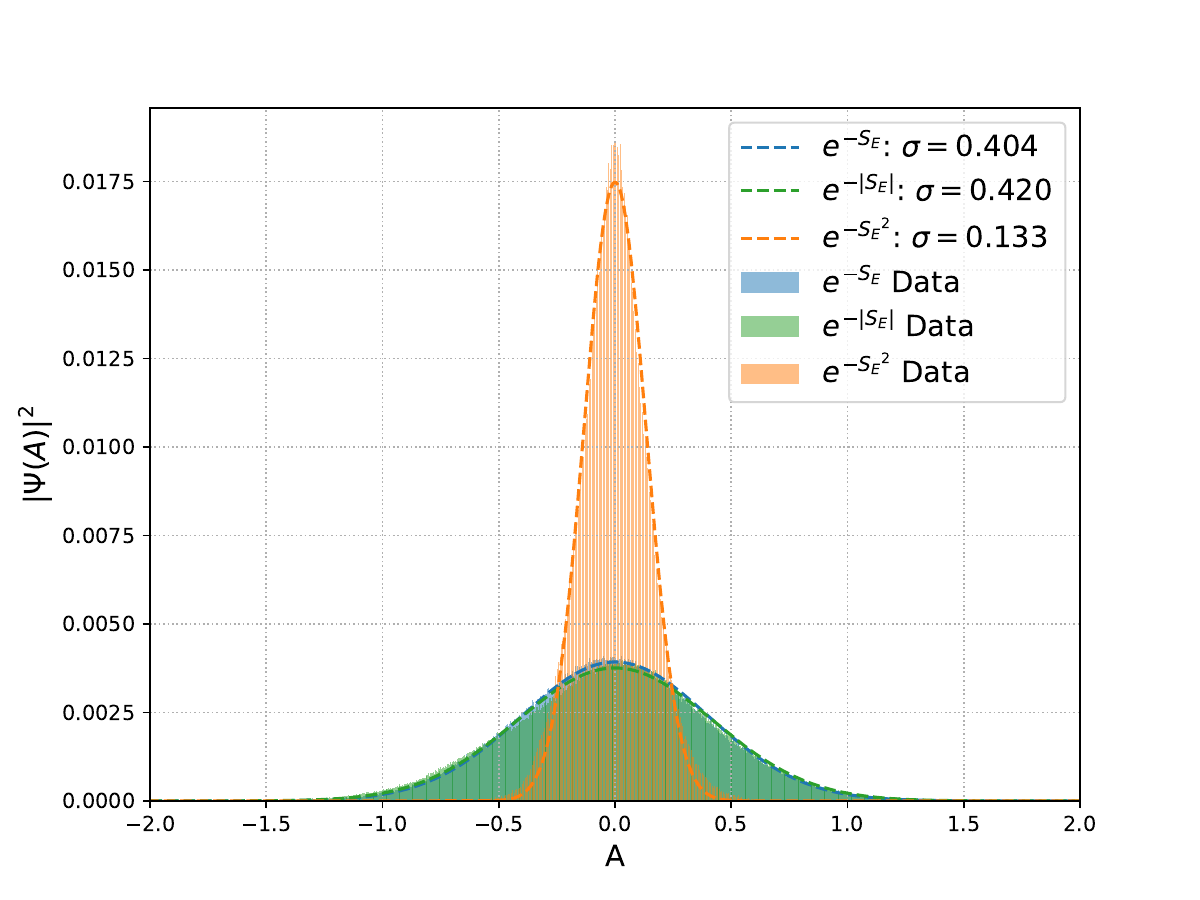}
    \caption{The `no-boundary' ground state wavefunctions (squared) for all three probability measures ($\Lambda = -1$).}
    \label{fig:measure_comparison09}
\end{figure}

\begin{figure} 
    \centering
    \includegraphics[width=0.8\linewidth]{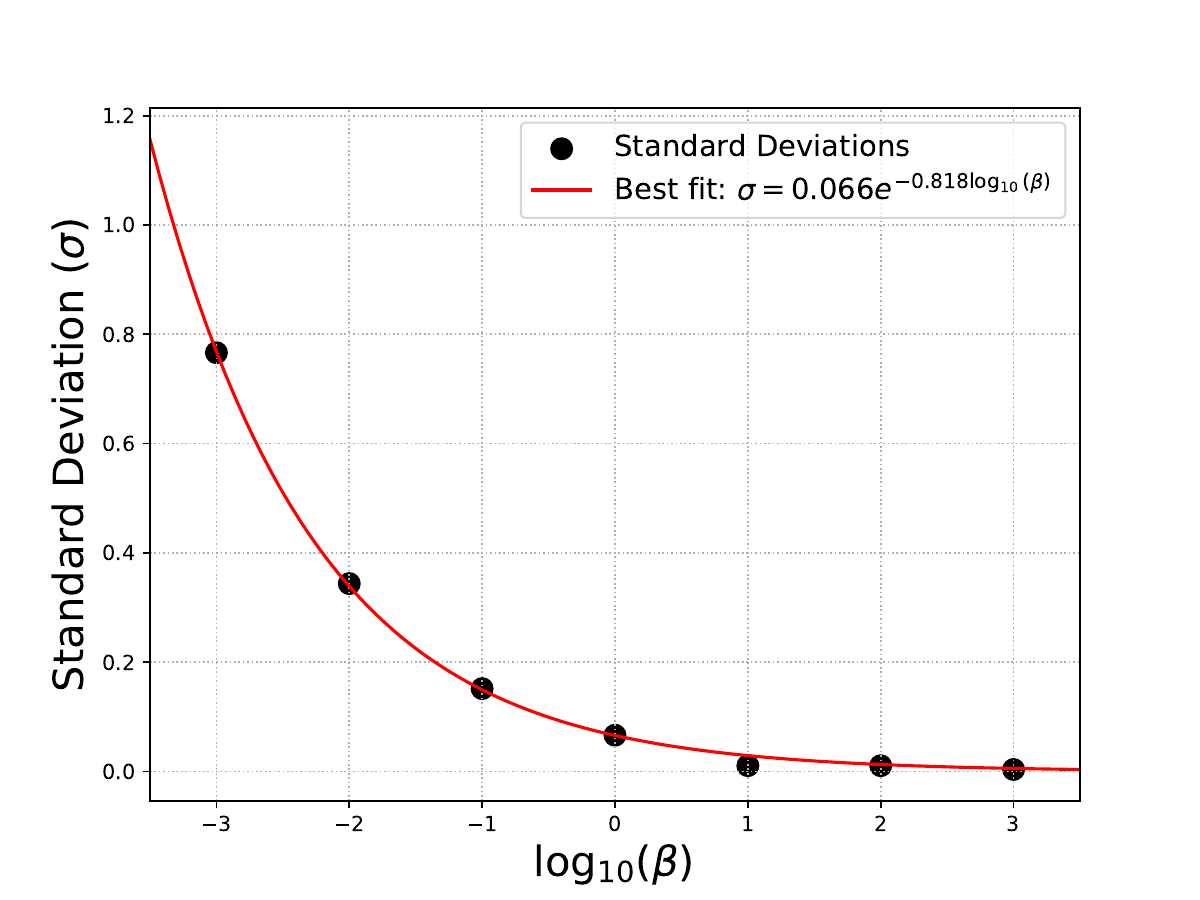}
    \caption{Finding an optimal value for $\beta$ from the wavefunctions ($\Lambda = -1$).}
    \label{fig:measure_comparison10}
\end{figure}

\begin{figure} 
    \centering
    \includegraphics[width=1\linewidth]{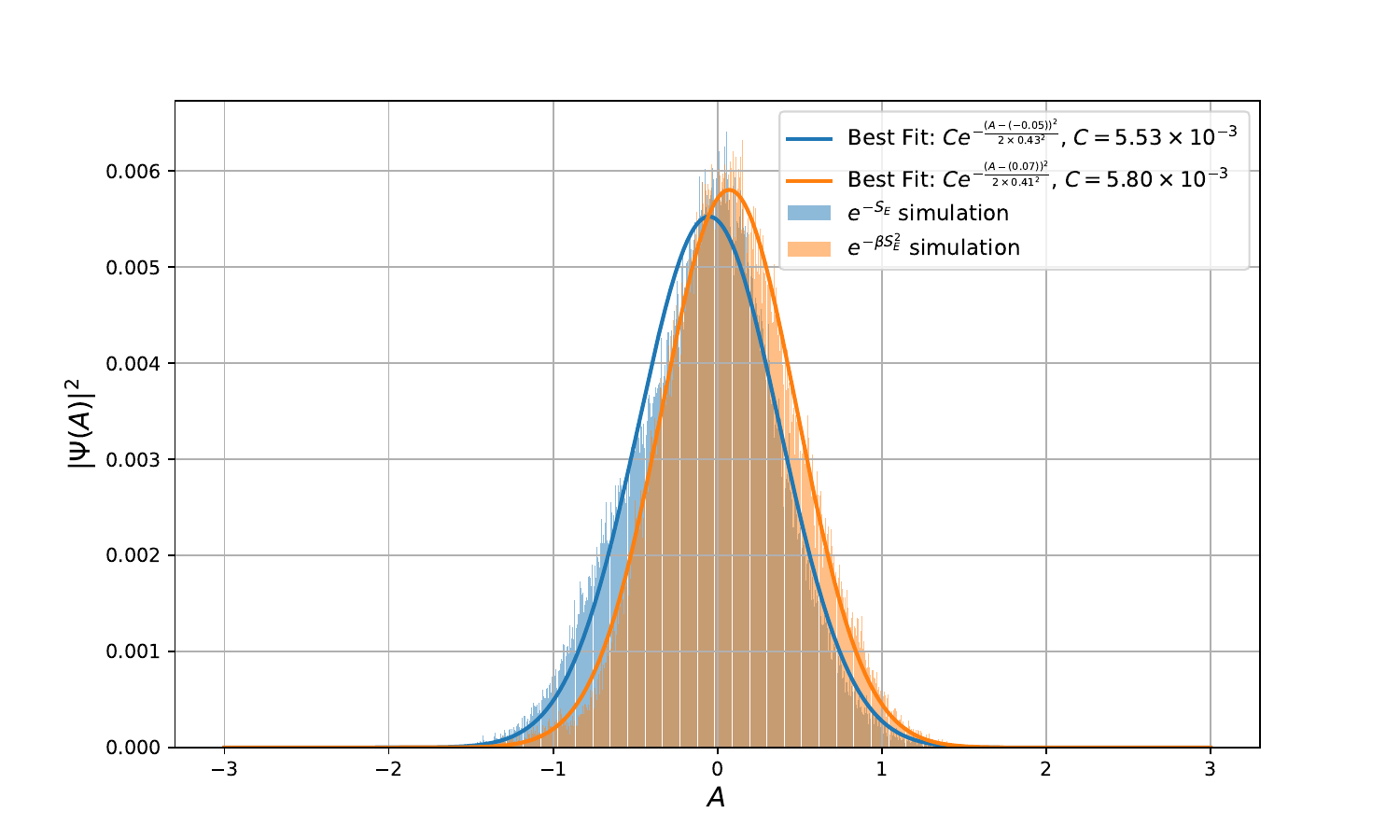}
    \caption{Comparison between the wavefunctions for the $e^{-S_E}$ and the $e^{-\beta_{\text{opt}} {S_E}^2}$ cases ($\Lambda$ = -1).}
    \label{fig:BetaStdComparisonInterpolated}
\end{figure}

It turns out that, as it was for the simple harmonic oscillator (Appendix \ref{sec-app}), the wavefunctions for the $e^{-S_E}$ and the $e^{-|S_E|}$ cases are a close match. This is demonstrated in Figure \ref{fig:measure_comparison09}. In contrast to this, the wavefunction for the $e^{-S_E^2}$ case is significantly `squished'. We repeat the procedure outlined in Appendix \ref{sec-app}, considering, instead of $e^{-S_E^2}$, the measure $e^{-\beta S_E^2}$, and calculating the optimum value for $\beta$ by matching the wavefunction with the standard case. The results of this matching procedure are shown in Figure \ref{fig:measure_comparison10} (where $\sigma$ is the standard deviation of the wavefunction), and the value of $\beta_{\text{opt}}$ is calculated to be,
\be
\label{bo-wave}
\beta_{\text{opt}} = 10^{-2.2148}.
\ee
The ground state wave function for this value is shown in Figure \ref{fig:BetaStdComparisonInterpolated}, along with the wavefunction for the $e^{-S_E}$ case (various parameter values were chosen to be, $\epsilon = 0.01$, $N = 1500$ (i.e., $T=15$), $N_{MC} = 10^{5}$, $\Delta = 0.2$ and $\bar n = 5$, with the initial path chosen randomly in the range [-5, 5]).

All of these wavefunctions appear to be Gaussians centered near $A=0$, and show that the quantum universe is most likely to have a very small size when in its ground state.

\begin{figure} 
    \centering
    \includegraphics[width=\linewidth]{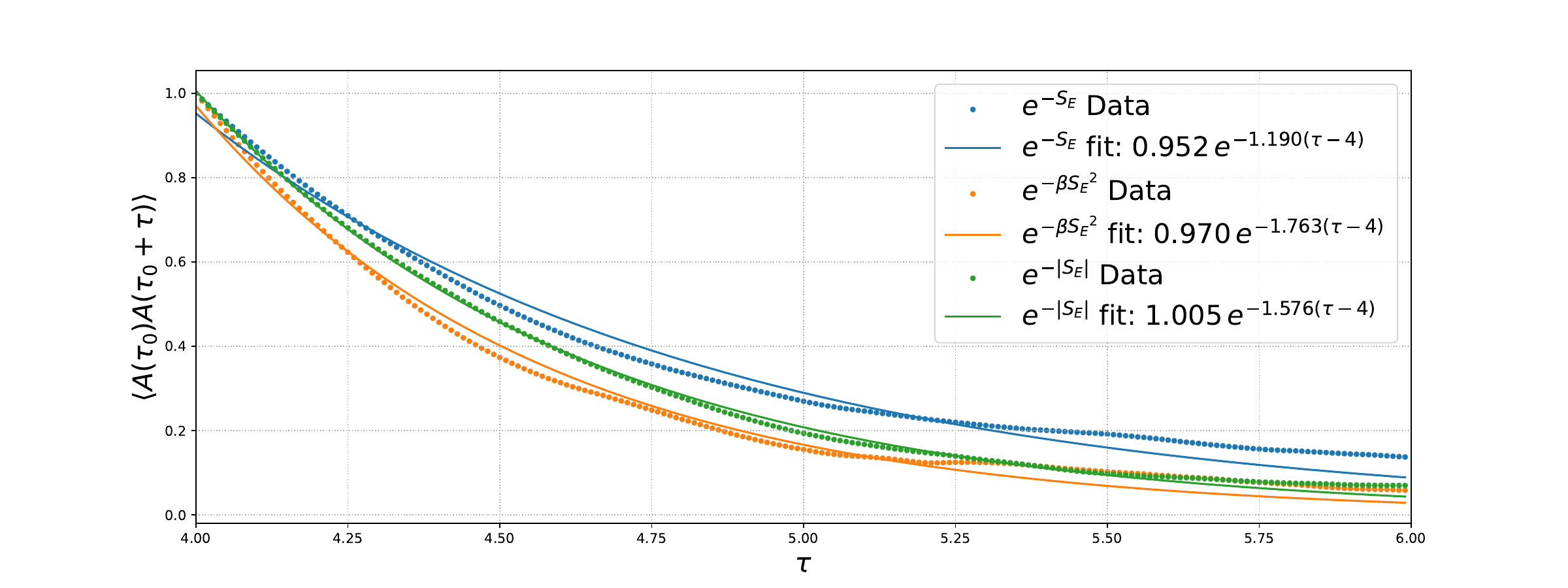}
    \caption{Two point correlation functions (along with the best fit functions) for all three probability measures ($\Lambda = -1$).}
    \label{fig:measure_comparison11freefit}
\end{figure}

Next, we proceed to computing the two point correlation function $\langle A(\tau_{0})A(\tau_0 + \tau) \rangle$ for all three probability measures. The results are shown in Figure \ref{fig:measure_comparison11freefit} (with $\tau_{0} = 4$). We observe an exponentially declining trend with the argument of the exponent being the key difference between the different probability measures. We note  that for the $e^{-|S_E|}$ measure, neighboring points are less correlated as compared to the standard $e^{-S_E}$ measure. The $e^{-\beta S_E^2}$ measure gives an even smaller correlation, even with the optimum value of $\beta$ (\ref{bo-wave}). This is also evident from the exponents in the best fit functions.

\begin{figure} 
    \centering
    \includegraphics[width=0.8\linewidth]{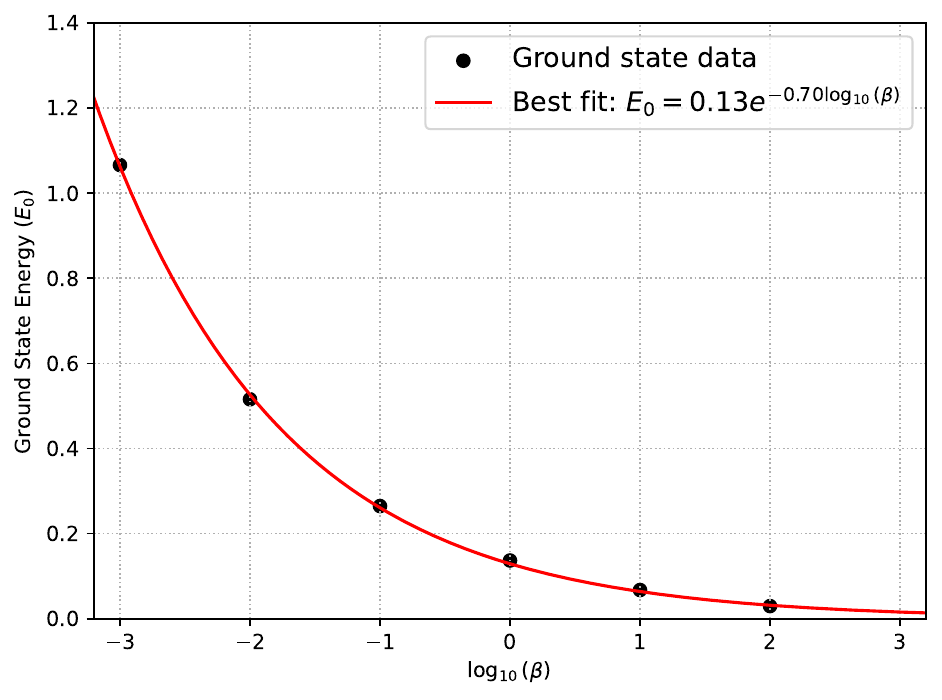}
    \caption{The ground state energy as a function of $\beta$, calculated using the $e^{-\beta S_E^2}$ probability measure ($\Lambda = -1$).}
    \label{fig:measure_comparison12}
\end{figure}

Lastly, we compute the ground state `energy' of the universe for all three probability measures,\footnote{Calculated using the method discussed in \cite{1981AnPhy.132..427C}, with the Virial theorem applied to our potential.} and as done for the simple harmonic oscillator, compute a value for $\beta_{\text{opt}}$. Various parameter values were chosen to be, $\epsilon = 0.1$, $N = 1000$ (i.e., $T = 100$), $N_{MC} = 10^5$, $\Delta = 0.2$ and $\bar{n} = 10$. The initial path in this case was chosen to be in the range [-100, 100]. The ground state energy as a function of $\beta$ is shown in Figure \ref{fig:measure_comparison12}, which gives (upon comparing with the ground state energy for the standard $e^{-S_E}$ case which is computed to be $E_0 = 1.0898$ -- for reference the value for the $e^{-|S_E|}$ case is $E_0 = 1.0924$),
\be
\beta_{\text{opt}} = 10^{-3.03745},
\ee
which is different from the value calculated by fitting to the ground state wavefunction (\ref{bo-wave}). (Equivalently, using the value of $\beta$ calculated before (\ref{bo-wave}), the ground state energy for the $e^{-\beta S_E^2}$ case turns out to be $E_0 \sim 0.6$.)

\subsection{Partly bounded case ($T < \pi/\sqrt{\Lambda}$)}

\begin{figure} 
    \centering
    \includegraphics[width=0.9\linewidth]{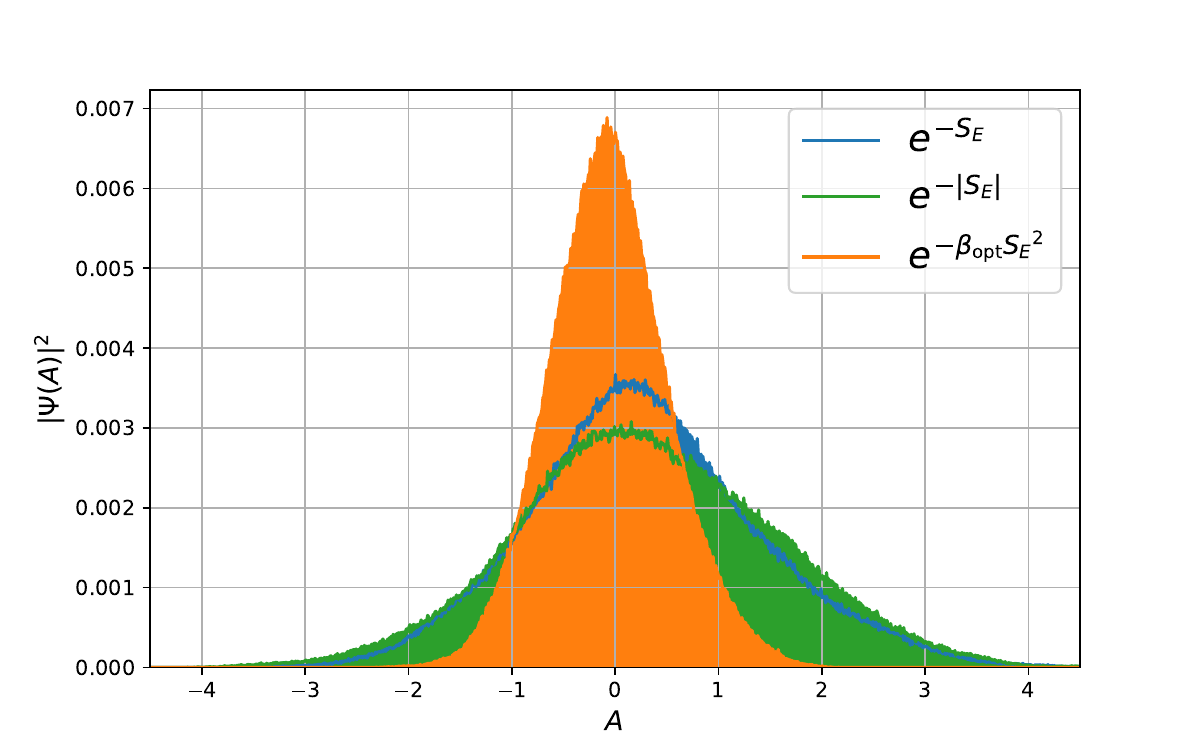}
    \caption{Ground state wavefunctions for the three probability measures ($\Lambda = 0.01$).}
    \label{fig:measure_comparison13}
\end{figure}

We now turn to an interesting case, where, despite a positive value of $\Lambda$, the Euclidean path integral is convergent given that the total time of integration $T$ satisfies $T < \pi/\sqrt{\Lambda}$ \cite{Carreau:1990is}. We choose $\Lambda = 0.01$, $\epsilon = 0.01$, and $N = 1000$ which gives $T = N\epsilon = 10$ clearly smaller than $\pi/\sqrt{\Lambda} \sim 31.4$. Other parameters were $N_{MC} = 10^6$, $\Delta = 0.2$, $\bar n = 5$, and $N_\text{therm} = 1.5 \times 10^6$. The initial path considered was in the range [-4.5, 4.5]. We again keep the initial point of the path fixed at $A(0) = 0$, and the results are computed by averaging over four independently thermalized arrays (i.e., four independent simulation runs). The resulting wavefunctions are shown in Figure \ref{fig:measure_comparison13}. The standard case reproduces the results of \cite{Ali:2018nql}, and the $e^{-|S_E|}$ case appears to be close to, but distinct from, the standard case (it has a larger tail on one end). The $e^{-\beta S_E^2}$ wavefunction is substantially different from the other two cases in that it is more squeezed around $A=0$. It might be interesting to note any deviations from Gaussianity (as evident in the $e^{-|S_E|}$ case, for example) by computing higher moments.

\subsection{Unbounded case ($\Lambda = 1$)}

Finally, we address the case where the Euclidean action is unbounded from below. In this case, a unique ground state does not exist. For each of the three choices we make: the standard measure (which is sampled by implementing a cutoff at $S_E = 0$), the $e^{-|S_E|}$ measure, or the $e^{-\beta S_E^2}$ measure, there is an infinite number of paths that minimize the action. In any given run, the MCMC process proceeds to the nearest minima and then stays around its neighborhood. This means that any computations will depend on the chosen random initial configuration. Thus, although we cannot compute unique ground state wavefunctions, or unique correlation functions, we can still draw generic conclusions about some observables. In particular, we compute the expectation value of the volume of the universe $\langle V(\tau) \rangle \sim \langle A^2(\tau) \rangle $ as a function of time $\tau$.

\begin{figure}[h!]
    \centering
    \begin{subfigure}[t]{0.5\textwidth}   
        \centering
        \includegraphics[width=\textwidth]{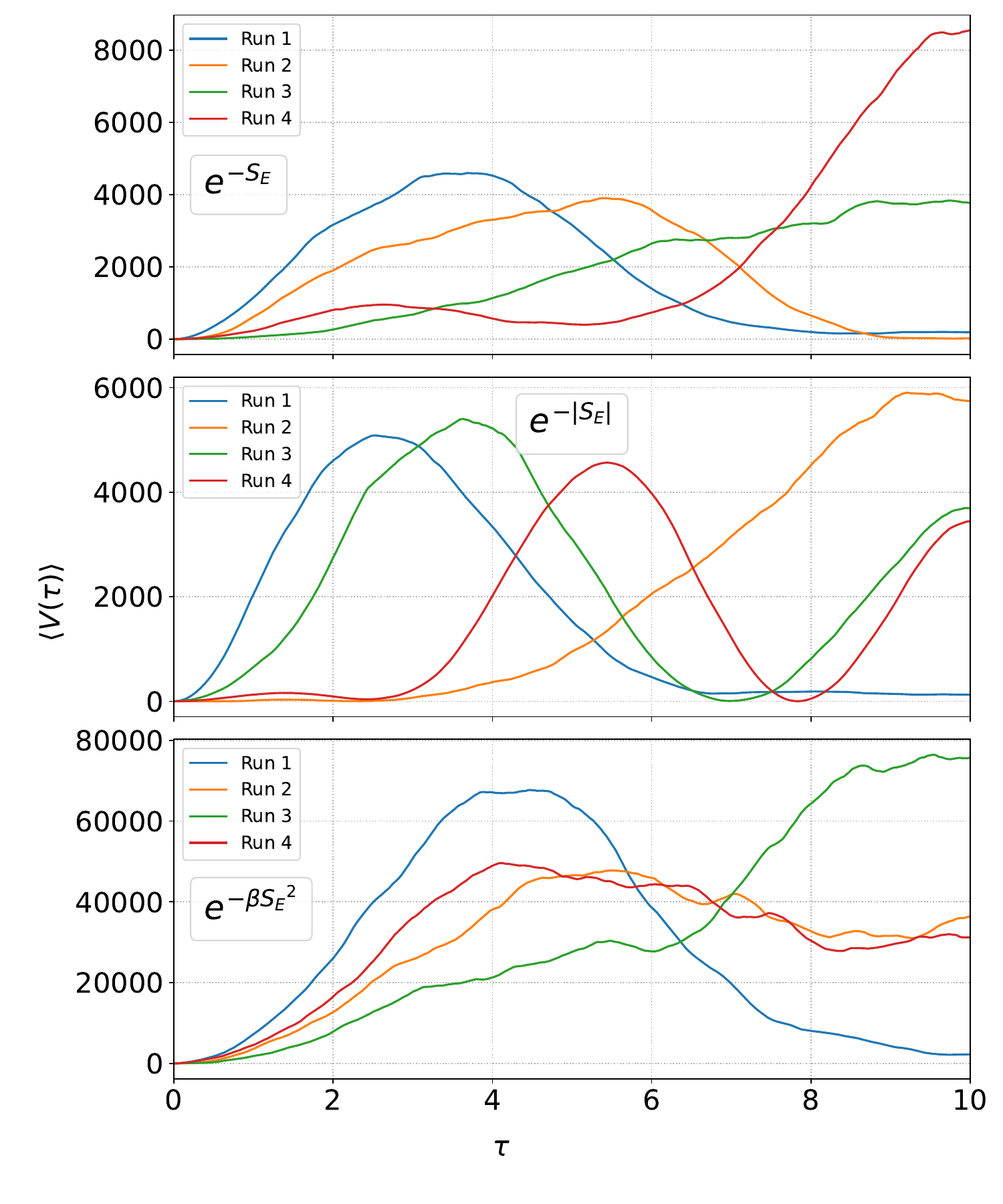}
        \caption{$T = 10$}
    \end{subfigure}
    \hspace{-0.015\textwidth}                 
    \begin{subfigure}[t]{0.5\textwidth}   
        \centering
        \includegraphics[width=\textwidth]{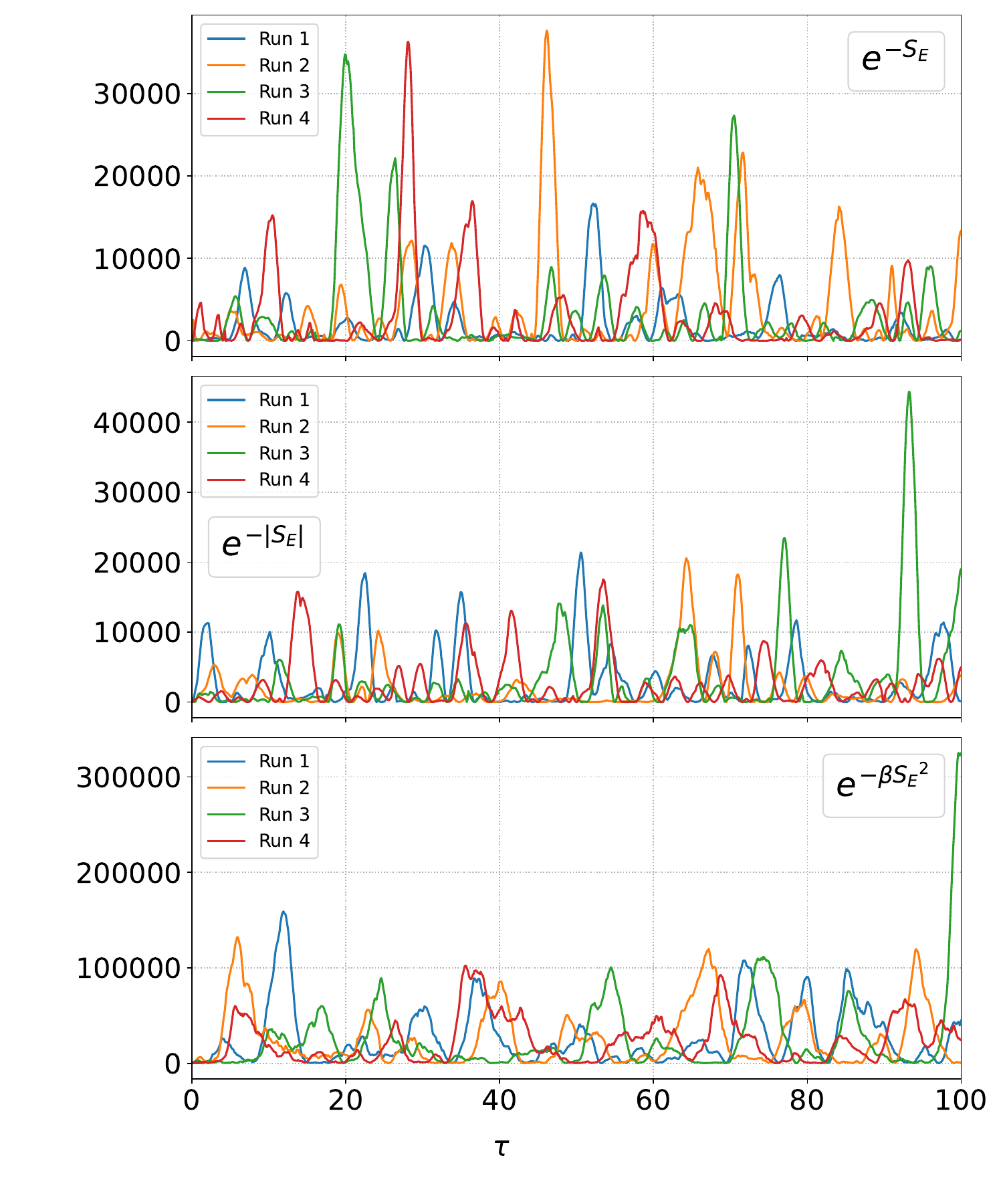}
        \caption{$T = 100$}
    \end{subfigure}
    \caption{Comparison of the expectation value of the volume of the universe $\langle V(\tau) \rangle$ for all three probability measures ($\Lambda = 1$). \textit{Left:} $T=10$. \textit{Right:} $T=100$. Each plot was sampled at 1000 values of $\tau$.}
    \label{fig:comparison_grid}
\end{figure}

We carried out four independent runs (with a different starting configuration) for each case. The value of $T$ is chosen to be arbitrarily larger than $\pi / \sqrt\Lambda$. We run simulations for two different values of $T$, first for $T = 10$ ($\epsilon = 0.01$, $N = 1000$), and then for $T = 100$ ($\epsilon = 0.1$, $N = 1000$) with $N_{MC} = 10^6$. Other parameter values were $\Delta = 0.2$, $\bar n = 5$, and we keep the initial point fixed at $A(0)=0$.\footnote{The value of $\beta$ used for the $e^{-\beta S_E^2}$ plots was a bit different different from the $\beta_{\text{opt}}$ we calculated above using the wavefunction. However, this should not affect the qualitative results, and the conclusions we draw from them.} The results are shown in Figure \ref{fig:comparison_grid}. We find that, for all three measures, there is a high probability of the universe to expand to a large size, starting from size zero. Furthermore, as we increase the total time $T$, the universe undergoes multiple cycles of expansion and contraction. One notable difference in the $e^{-\beta S_E^2}$ case is that the universe reaches much larger volumes. This fact can be traced back to the small value of $\beta$ (\ref{bo-wave}), which produces a widely spread sampling distribution.

\section{Summary} \label{sec-sum}

We have presented here a Euclidean quantum cosmological model for a homogeneous and isotropic FLRW universe with a positive scalar curvature, and a cosmological constant term. The theory is studied in the `dust-time gauge' where a pressureless dust field is used as an internal clock. Since this model is not analytically tractable, we numerically evaluate the (Euclidean) path integral and compute various quantities like the wavefunction of the universe, correlation functions, and mean volume.

Given that the gravitational Euclidean action is unbounded from below, the Euclidean path integral is not well defined. In \cite{Ali:2018nql}, this was overcome by selecting only a subset of paths in the path integral for which $S_E = 0$. Here we have proposed a different approach in which the measure of the (Euclidean) path integral is modified from $e^{-S_E}$ to either $e^{-|S_E|}$ or $e^{-\beta S_E^2}$. This is to be viewed as a definition for a Euclidean theory of quantum cosmology. (A detailed analysis of these measures for the quantum harmonic oscillator -- including how we fixed a value for $\beta$ -- appears in the Appendix.)

With these modified measures, we computed the path integral for three different choices of the cosmological constant: $\Lambda = -1$ (where the action is bounded below), $\Lambda < \pi^2/T^2$ (where the action is unbounded, but the path integral exists), and $\Lambda = 1$ (where the action is unbounded). For the first two choices, we computed the (unique) ground state wavefunctions (along with the two point correlation function, and the ground state energy of the universe for the first choice). We found that the $e^{-|S_E|}$ case is similar to the standard measure (with some differences), whereas, the $e^{-\beta S_E^2}$ case is substantially different.

For the unbounded case, a unique ground state does not exist, and we calculated the expected value of the volume of the universe. We found that there is dependence on the initial chosen configuration, and that all three measures give qualitatively similar results, with the universe expanding to a large size (starting from zero size). For larger values of the total time $T$, the universe undergoes multiple expansion and contraction cycles, and reaches a much larger size with the $e^{-\beta S_E^2}$ measure as compared to the other two measures.

Our results seem to indicate some theory dependence for the parameter $\beta$. A useful generalization might be to consider regulating the path integral with measures of the form $e^{-\beta(S_E) S_E}$, where $\beta(S_E)$ is now a function of the Euclidean action. The two cases we have considered here correspond to $\beta(S_E) = \text{sgn}(S_E)$, and $\beta(S_E) = S_E$ respectively. We hope to report on this in the future. There are other interesting proposals as well for stabilizing theories with actions unbounded from below \cite{Greensite:1983yc, Myers:1991yq}. It would be interesting to see what these methods produce for the cosmological scenario considered here.

\appendix

\section{\label{sec-app} Simple Harmonic Oscillator}

The discretized Euclidean action for the simple harmonic oscillator takes the form,
\be
\label{osc-act}
S = \epsilon \sum_{i=1}^{N} \left( \frac{1}{2} m \frac{(x_{i+1} - x_i)^2}{\epsilon^2} + \frac{1}{2} {\mu}^2 {x_i}^2 \right),
\ee
where $m$ is its mass, $\omega$ is its angular frequency, and $\mu^2 = m \omega^2$.

\subsection{Standard case}

We first describe the standard case (path weight $ = \exp(-S_E)$), which serves both as a guide towards the modified cases, and as a useful numerical check of our code.

\begin{figure} 
    \centering
    \includegraphics[width=0.8\linewidth]{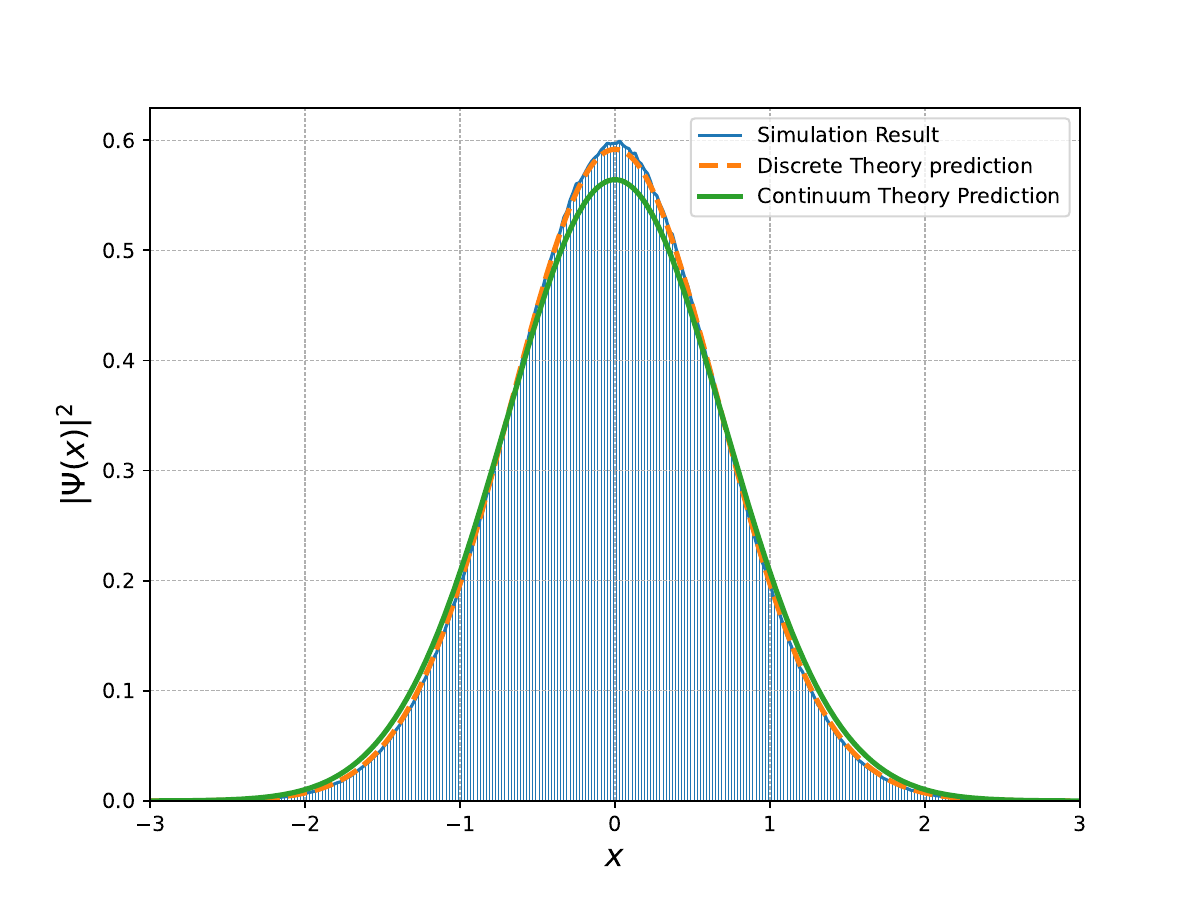}
    \caption{Comparison between simulation results and analytical wavefunctions for the quantum harmonic oscillator ground state for both the continuum and discretized theories. A full treatment of the latter is given in \cite{1981AnPhy.132..427C}.}
    \label{fig:measure_comparison00}
\end{figure}

The ground state wavefunction is computed by generating a sample path at each Monte Carlo iteration, and binning the positions at different times into regular intervals of size $ = 0.0025$. Various other parameter values were chosen to be $\epsilon = 0.1, N = 1000$, $N_{MC} = 10^4$, $\Delta = 1.5$ and $\bar{n} = 10$. Since the analytical solution is available, we can compare our simulation results with the analytical result. This is shown in Figure \ref{fig:measure_comparison00}, which also shows the analytical result of the discretized theory. Note that, given a finite discretization scale, the discrete theory results are slightly different from the continuum results.

\begin{figure} 
    \centering
    \includegraphics[width=0.8\linewidth]{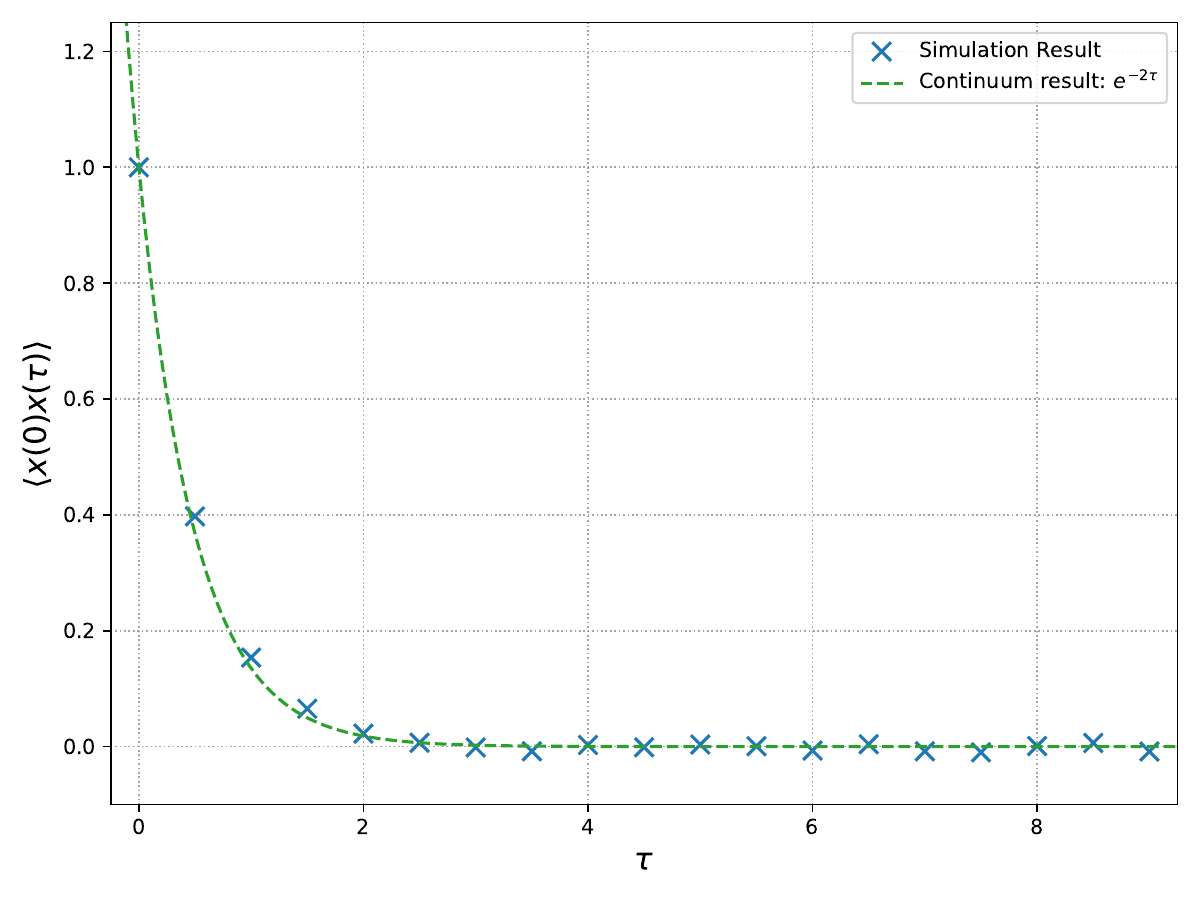}
   \caption{Comparison between the simulation result and the analytically calculated correlation function for the quantum harmonic oscillator ground state.}
    \label{fig:measure_comparison01}
\end{figure}

Next, we numerically compute the two point correlation function $\langle x(0)x(\tau) \rangle$, which is analytically computed to be,
\be
\langle x(0)x(\tau) \rangle = \frac{1}{2m\omega}e^{-i\omega \tau},
\ee
or, after Wick rotating (and plugging in parameter values $m = 0.5$, $\mu^2 = 2$),
\be
\label{osc-corr}
\langle x(0)x(\tau) \rangle = \frac{1}{2}e^{-2\tau}.
\ee
We plotted this analytical result against the data points obtained from the correlation simulations (for $\epsilon = 0.5$, $N =1000$, $N_{MC} = 10^4$, $\Delta = 1.5$ and $\bar{n} = 10$) shown in Figure \ref{fig:measure_comparison01}.

\subsection{Modified Euclidean action}

We now proceed to test two alternatives for the probability measure: $e^{-|S_E|}$ and $e^{-S_E^2}$. The simulations were performed using the same parameters as before ($\epsilon = 0.5$, $N = 1000$, $N_{MC} = 10^4$, $\Delta = 1.5$ and $\bar{n} = 10$).

\begin{figure} 
    \centering
    \includegraphics[width=0.8\linewidth]{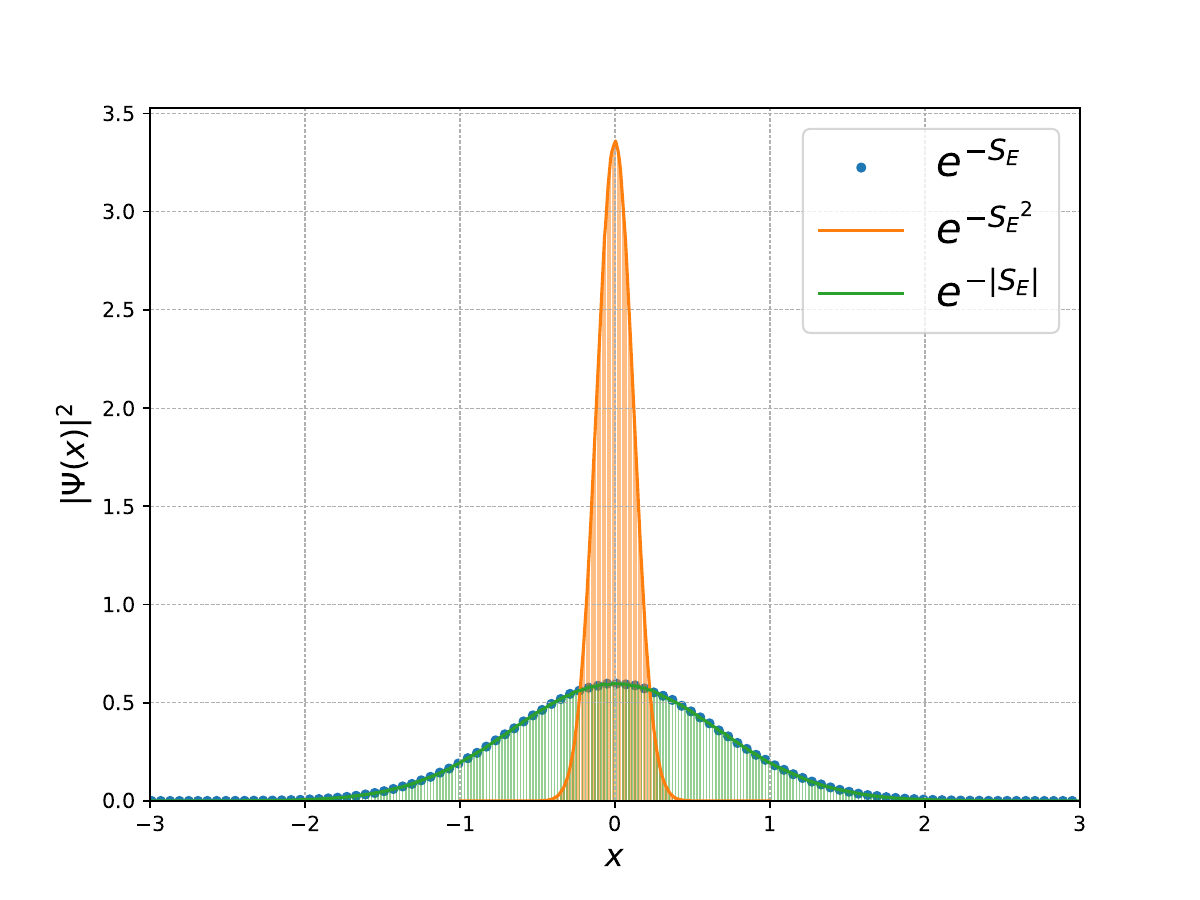}
    \caption{Comparison of the oscillator ground state wavefunction for the probability measures $e^{-|S_E|}$ and $e^{-S_E^2}$.}
    \label{fig:measure_comparison02}
\end{figure}

Figure \ref{fig:measure_comparison02} illustrates the comparison of the ground state wavefunctions obtained using the two new probability measures. It is quite evident that the $e^{-|S_E|}$ measure has no significant effect on the results of our simulations. This is to be expected, since the Euclidean action for the oscillator (\ref{osc-act}) is always positive. On the other hand, we note that the $e^{-S_E^2}$ measure has a significant impact on the resulting wavefunction: the standard deviation is reduced, resulting in a `squishing' of the wavefunction.

\begin{figure} 
    \centering
    \includegraphics[width=0.8\linewidth]{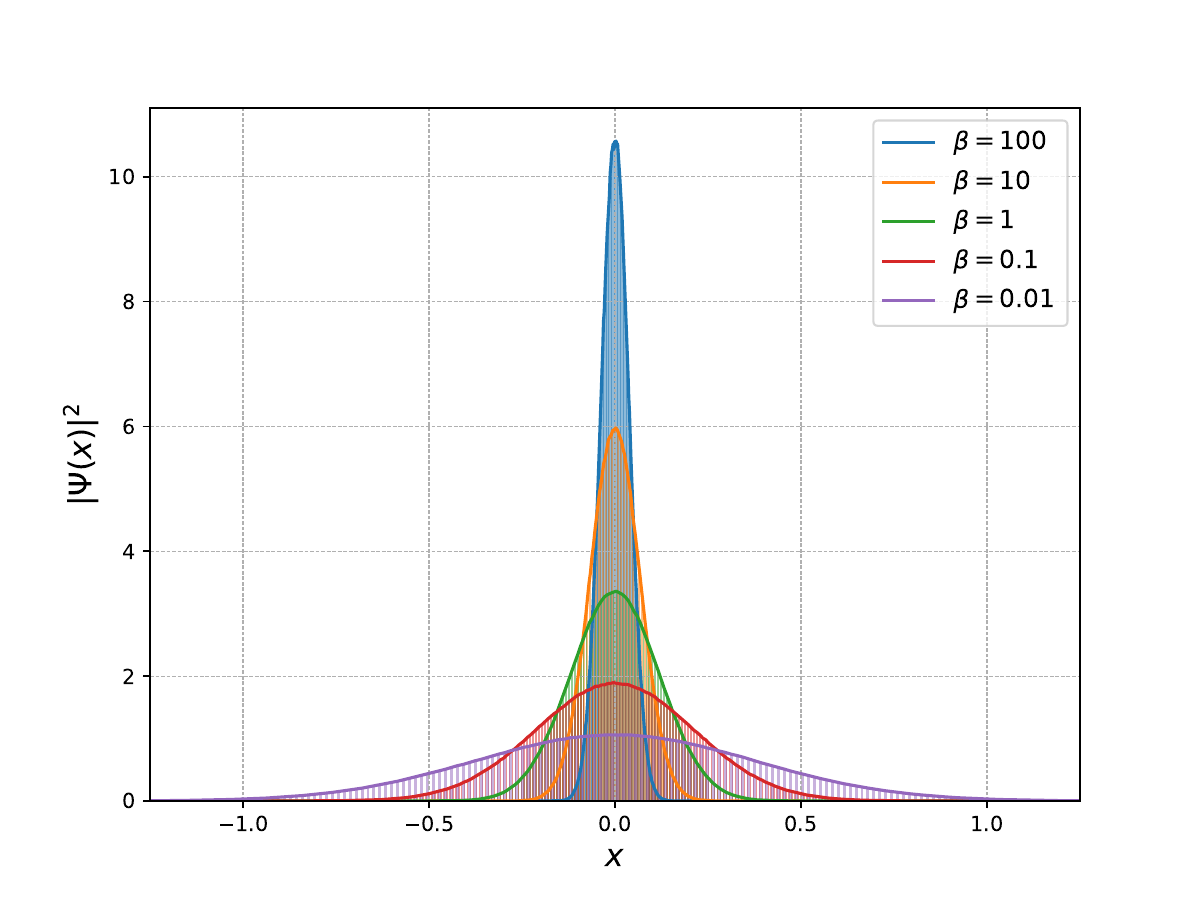}
    \caption{Effect of varying $\beta$ on the resulting wavefunction. It can be seen that larger values reduce the standard deviation, while smaller values increase it.}
    \label{fig:measure_comparison03}
\end{figure}

In order to make the $e^{-S_E^2}$ wavefunction fit the standard ground state wavefunction, one might consider introducing a parameter $\beta$ into the exponential. That is, we consider the probability measure $e^{-\beta S_E^2}$. In order to do this, we first consider the effect of varying this new parameter $\beta$ on the resulting wavefunction. Figure \ref{fig:measure_comparison03} shows the effects on the wavefunction by varying $\beta$. Evidently increasing this new parameter $\beta$ has the effect of squishing the wavefunction even further, while reducing it has the effect of spreading it out.

\begin{figure} 
    \centering
    \includegraphics[width=0.8\linewidth]{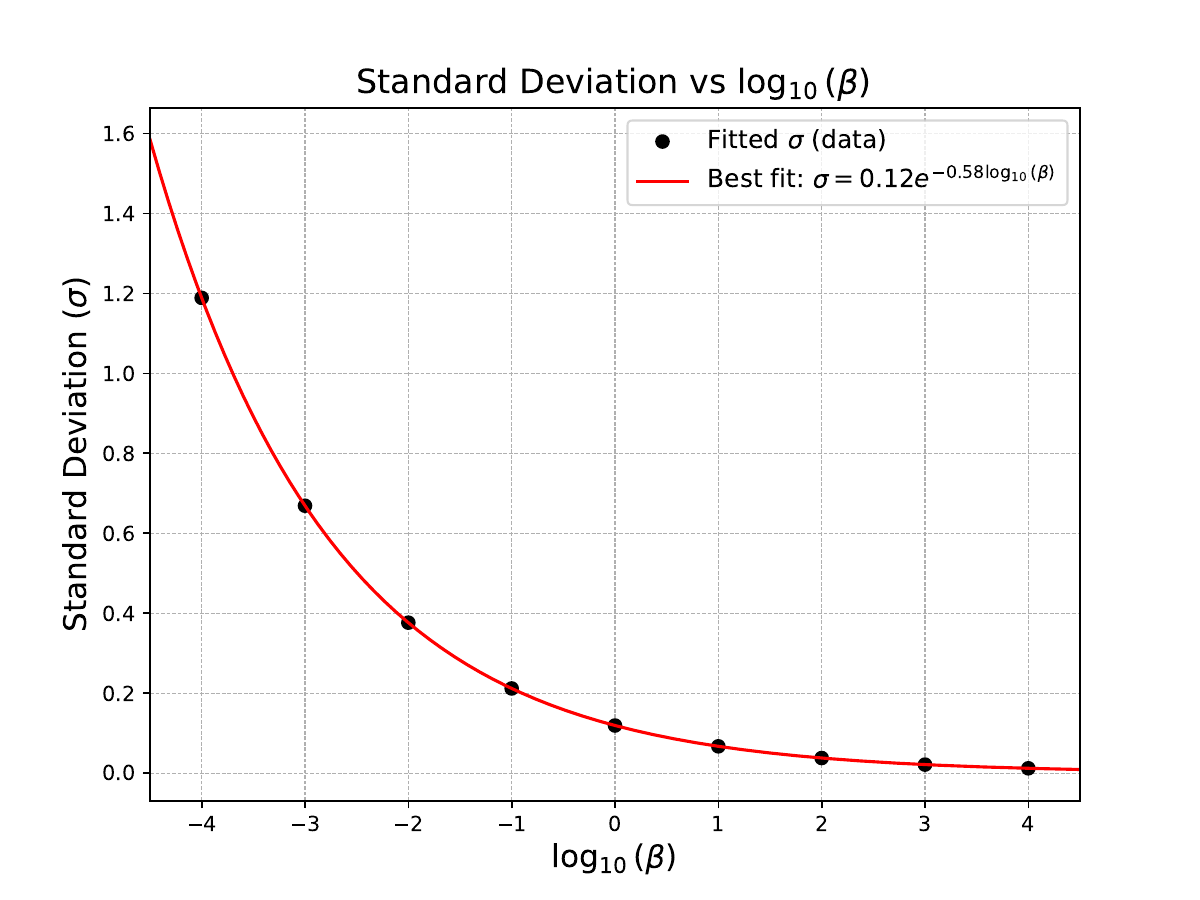}
    \caption{Fitting function for the dependence of the standard deviation $\sigma$ of the wavefunction on the parameter $\beta$.}
    \label{fig:measure_comparison04}
\end{figure}

In order to figure out the optimal value of $\beta$ required to fit to the standard wavefunction, we note that the ground state wavefunction is a Gaussian, and changing $\beta$ changes its standard deviation $\sigma$. We numerically infer this relationship by simulating the wavefunction for different values of $\beta$, fitting Gaussian functions to the resulting wavefunctions, plotting the standard deviations from these fits as a function of $\beta$, and then using a fitting function to find the relationship. The results are shown in Figure \ref{fig:measure_comparison04}, which show that over a large range of values of $\beta$ (from $10^{-4}$ to $10^4$), the following exponential decay fits the data neatly,
\be
\label{sig-beta}
\sigma = 0.12 ~e^{-0.58 ~\text{log}_{10}(\beta)}.
\ee

\begin{figure} 
    \centering
    \includegraphics[width=0.8\linewidth]{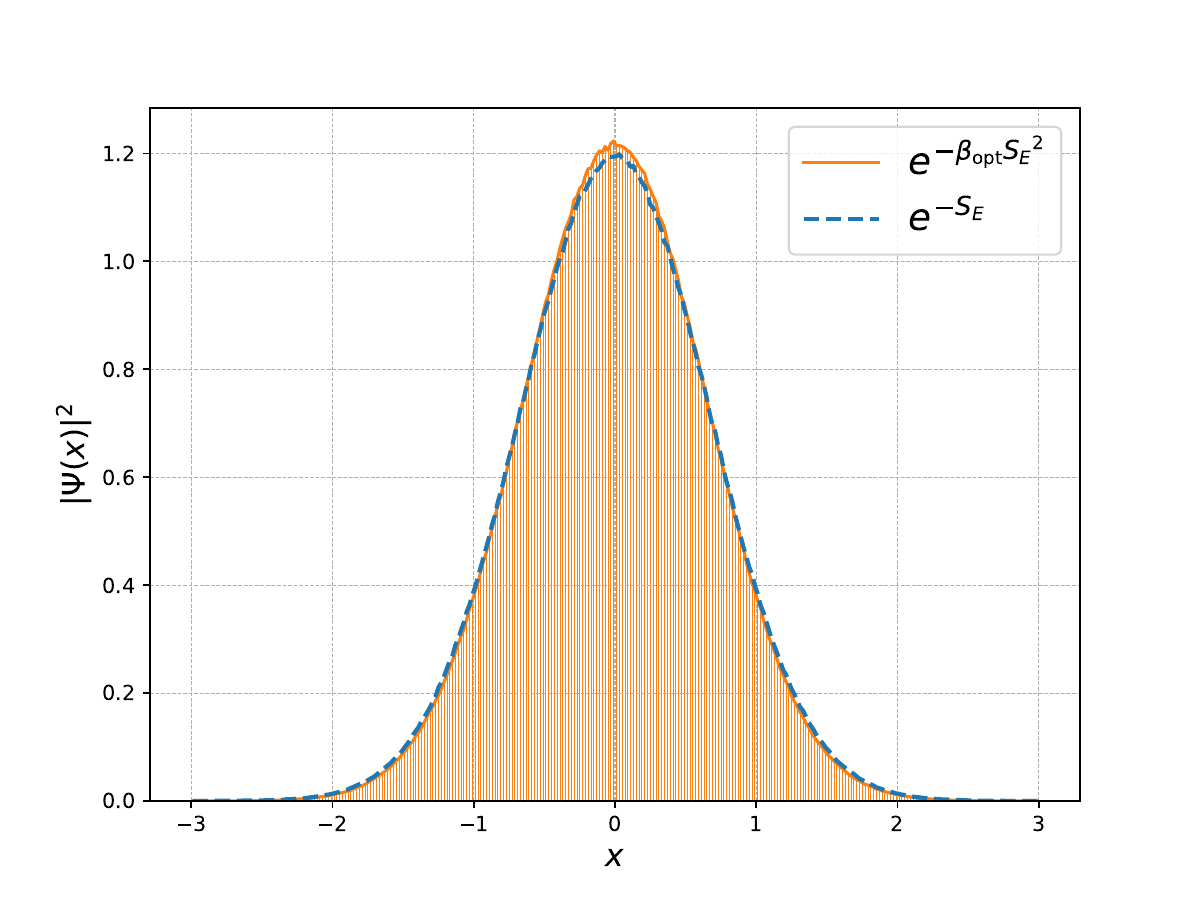}
    \caption{The ground state wavefunction for the measure $e^{-\beta_{\text{opt}}S_E^2}$ (blue dotted line). The orange solid line is the standard wavefunction for the measure $e^{-S_E}$.}
    \label{fig:measure_comparison05}
\end{figure}

The standard deviation of the (numerical) ground state wavefunction calculated using the usual probability distribution ($e^{-S_E}$) comes out to be $\sigma = 0.6687$. Substituting this value into (\ref{sig-beta}), we obtain 

\be
\label{bo-sho}
\beta_{\text{opt}} = 10^{-2.9618}
\ee

as our optimal choice. Repeating the MCMC simulation for the original parameter values using the measure $e^{-\beta_{\text{opt}}S_E^2}$, we obtain the wavefunction shown in Figure \ref{fig:measure_comparison05}, which is plotted against the wavefunction from the $e^{-S_E}$ case, and shows a close fit.

\begin{figure} 
    \centering
    \includegraphics[width=0.9\linewidth]{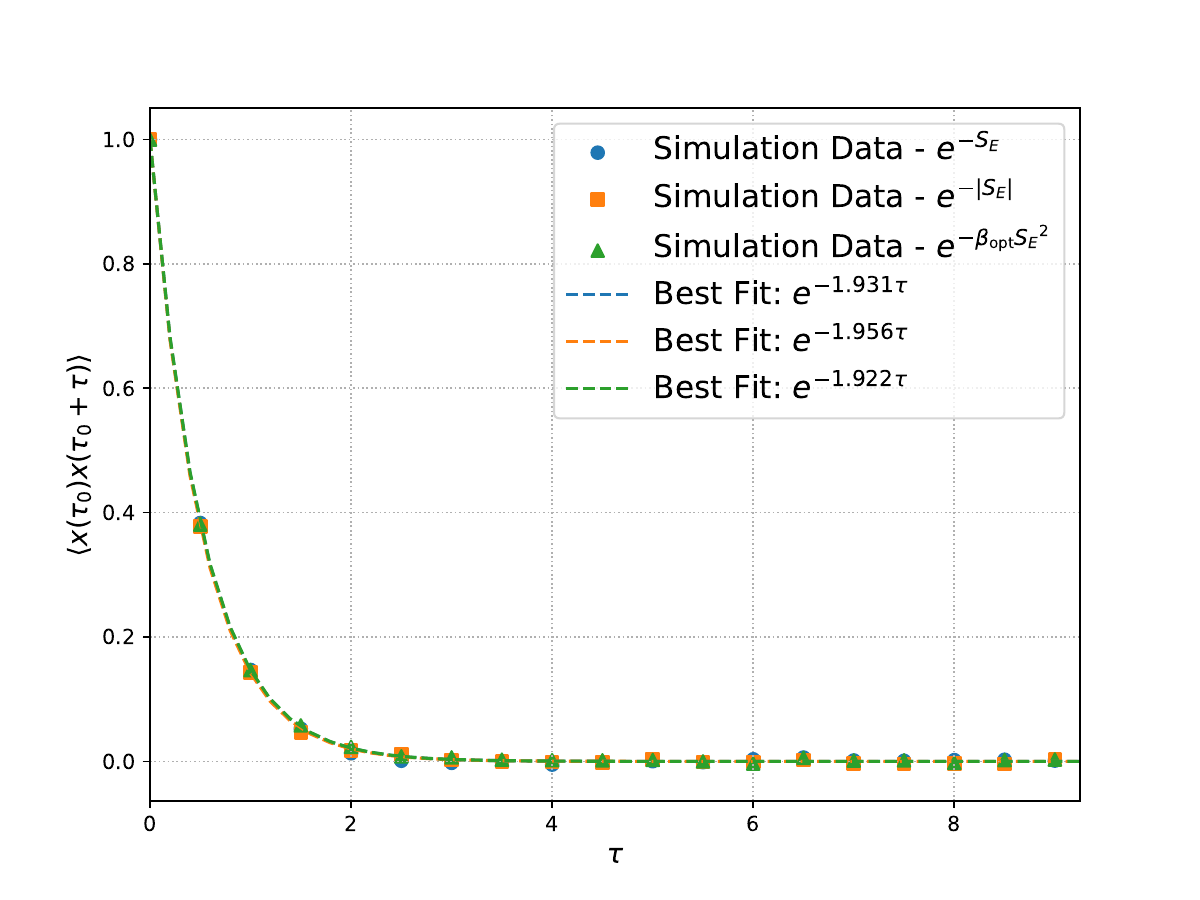}
    \caption{Plot of the two-point correlation function $\langle x(\tau_0)x(\tau_0 + \tau) \rangle$ for all three sampling methods. Scatter plot indicates the data points, while the dotted lines are functions fitted to the data points.}
    \label{fig:measure_comparison06}
\end{figure}

Another metric for testing the validity of the chosen $\beta_{\text{opt}}$ value is to compare the two point correlation functions $\langle x(\tau_{0})x(\tau_{0}+\tau) \rangle$ for the various sampling methods. Comparison between the two point correlation functions for all three sampling methods ($e^{-S_E}$, $e^{-|S_E|}$, and $e^{-\beta_{\text{opt}}S_E^2}$) is shown in Figure \ref{fig:measure_comparison06} (with $\tau_{0} = 0$). The correlation data points have been normalized so that the value $\langle x^2(\tau_{0}) \rangle$ is equal to 1. Plotted along with the data points are best fit interpolation lines of the form $e^{-b \tau}$, where $b$ is some fitting parameter. Evidently all three sampling methods give roughly the same results, with the values for $b$ being quite close to the analytically predicted value of 2 (\ref{osc-corr}).

\begin{figure} 
    \centering
    \includegraphics[width=0.9\linewidth]{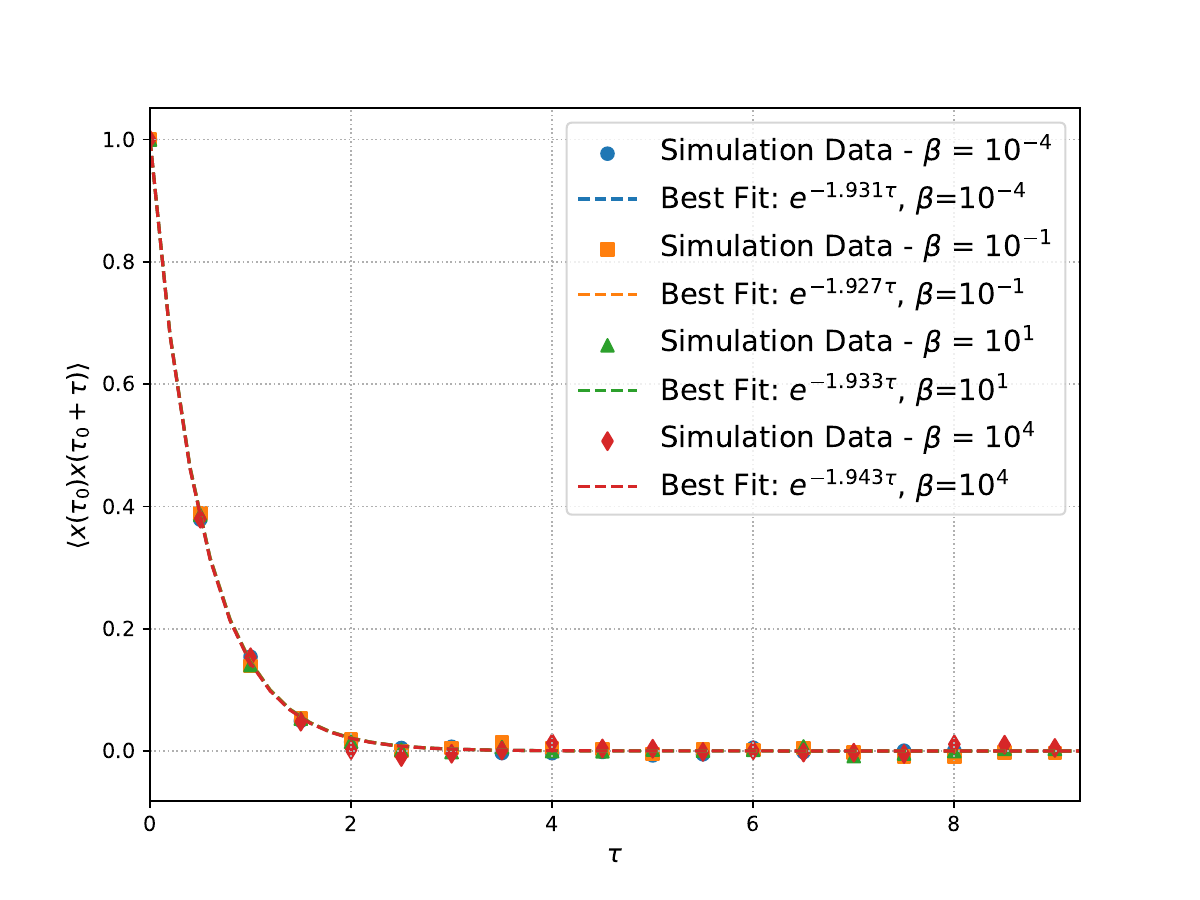}
    \caption{Plot of the two-point correlation function $\langle x(\tau_{0})x(\tau_0 + \tau) \rangle$ for the probability measure $e^{-\beta S_E^2}$ for various values of $\beta$.}
    \label{fig:measure_comparison07}
\end{figure}

The correlation function for the $e^{-\beta_{\text{opt}} S_E^2}$ case should be compared with the correlation functions obtained for various other values of $\beta$, shown in Figure \ref{fig:measure_comparison07} (with their corresponding best fit functions). Apparently, varying $\beta$ does not have any statistically significant bearing on the analytical form of the correlation function for the oscillator.

\begin{figure} 
    \centering
    \includegraphics[width=0.8\linewidth]{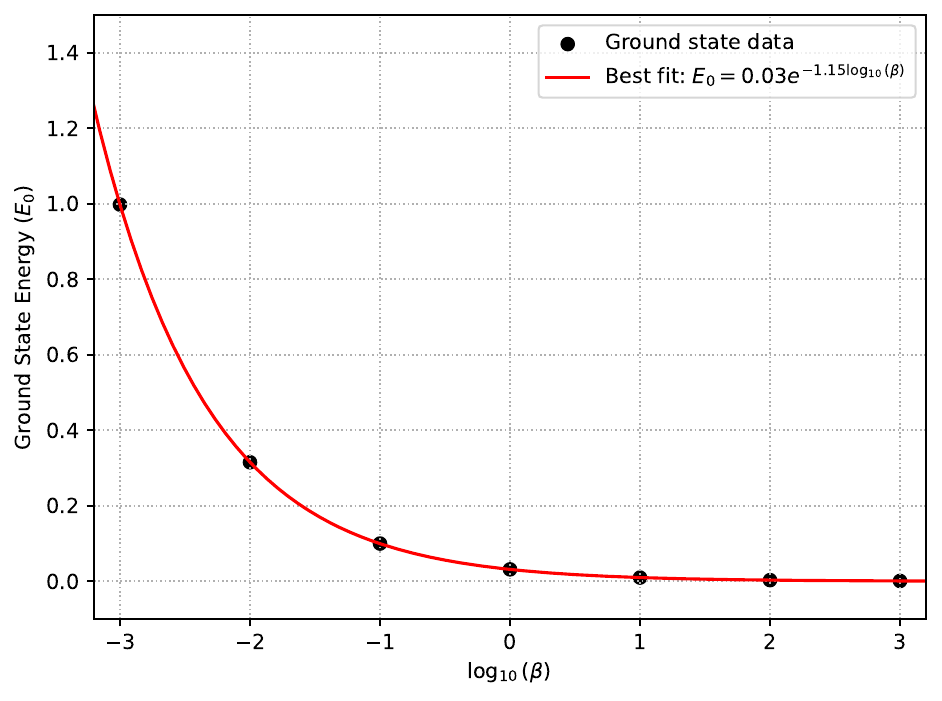}
    \caption{The best fit curve for the data points obtained upon calculating the ground state energy using the $e^{-\beta S_E^2}$ measure with different values of $\beta$.}
    \label{fig:measure_comparison08}
\end{figure}

Lastly, since the ground state energy of the harmonic oscillator is well-known, and can be calculated numerically using MCMC \cite{1981AnPhy.132..427C}, we calculated the ground state energy for the $e^{-\beta S_E^2}$ case for different values of $\beta$ and then used a best fit curve to find a value of $\beta$ that reproduces the usual result. This is shown in Figure \ref{fig:measure_comparison08} (with parameter values $\epsilon = 0.1$, $N = 1000$, $N_{MC} = 10^5$, $\Delta = 0.2$, and $\bar{n} = 10$) which gives a best fit equation,
\be
E_0 = 0.03 ~e^{{-1.15 ~\log_{{10}} (\beta)}},
\ee
and, an optimal value for $\beta_{\text{opt}}$ as,
\be
\beta_{\text{opt}} = 10^{-3.04832}
\ee
(for comparison, the ground state energies obtained using the $e^{-S_E}$ case, and the $e^{-|S_E|}$ case were $0.99901$ and $1.00121$ respectively, close to the analytical value of $E_0 = \omega/2 = 1$). This is quite close to the value of $\beta_{\text{opt}}$ calculated independently by fitting the standard deviation of the ground state wavefunction (\ref{bo-sho}) (the exponents are within 3\% of each other).

\bibliography{EuclideanQCbib}

\end{document}